\documentclass[reprint,
showpacs,preprintnumbers,
 amsmath,amssymb,
 aps,
prb,
floatfix,
]{revtex4-1}

\usepackage{graphicx}
\usepackage{dcolumn}
\usepackage{bm}

\newcommand{\lamno}{LaMn$_3$Mn$_4$O$_{12}$}
\newcommand{\bimno}{BiMn$_3$Mn$_4$O$_{12}$}
\newcommand{\namno}{NaMn$_3$Mn$_4$O$_{12}$}
\newcommand{\QP}{$AA'_3B_4$O$_{12}$}

\begin{document}

\preprint{APS/123-QED}

\title{Internal-strain mediated coupling \\between polar Bi and magnetic Mn ions \\in the defect-free quadruple-perovskite \bimno}

\author{Andrea Gauzzi}
\email{andrea.gauzzi@upmc.fr}
\author{Gwenaelle Rousse}
\affiliation{Institut de Min\'eralogie et de Physique des Milieux Condens\'es,
Universit\'e Pierre et Marie Curie-Sorbonne Universit\'es and CNRS, 75005 Paris, France}

\author{Francesco Mezzadri}
\author{Gianluca Calestani}
\affiliation{Dipartimento di Chimica - GIAF, Universit\`a degli Studi di Parma, 43100 Parma, Italy}

\author{Gilles Andr\'e}
\author{Fran\c{c}oise Bour\'ee}
\affiliation{Laboratoire L\'eon Brillouin, CEA-CNRS, 91191 Gif-sur-Yvette, France}

\author{Marco Calicchio}
\author{Edi Gilioli}
\author{Riccardo Cabassi}
\author{Fulvio Bolzoni}
\author{Andrea Prodi}
\affiliation{Istituto dei Materiali per Elettronica e Magnetismo\\
Consiglio Nazionale delle Ricerche, Area delle Scienze, 43100 Parma, Italy}

\author{Pierre Bordet}
\affiliation{Institut N\'eel-CNRS, 25, rue des Martyrs, 38042 Grenoble, France}
\author{Massimo Marezio}
\affiliation{CRETA-CNRS, 25, rue des Martyrs, 38042 Grenoble, France }

\date{\today}

\begin{abstract}
By means of neutron powder diffraction measurements in the 1.5-300 K range, we investigated the effect of the polar Bi$^{3+}$ ion on the magnetic ordering of the Mn$^{3+}$ ions in \bimno, the counterpart with \textit{quadruple} perovskite structure of the \textit{simple} perovskite BiMnO$_3$. At all temperatures, the data are consistent with a \textit{noncentrosymmetric} spacegroup $Im$ which contrasts the \textit{centrosymmetric} one $I2/m$ previously reported for the isovalent and isomorphic compound \lamno. This difference gives evidence of a Bi$^{3+}$-induced polarization of the lattice in \bimno. At low temperature, the two Mn$^{3+}$ sublattices of the $A'$ and $B$ sites are found to order antiferromagnetically (AFM) in an independent fashion at 25 and 55 K, respectively, similarly to the case of \lamno. However, both magnetic structures of \bimno~ radically differ from those of \lamno. Specifically, in \bimno~ the moments $\textbf{M}_{A'}$ of the $A'$ sites form an anti-body AFM structure, whilst the moments \textbf{M}$_{B}$ of the $B$ sites result from a large and \textit{uniform} modulation $\pm \textbf{M}_{B,b}$ along the $b$-axis of the moments \textbf{M}$_{B,ac}$ in the $ac$-plane of an $E$-type structure. The modulation is found to be strikingly correlated with the displacements of the Mn$^{3+}$ ions induced by the Bi$^{3+}$ ions. Our symmetry analysis of this correlation unveils a strong magnetoelastic coupling between the internal strain field created by the Bi$^{3+}$ ions and the moment of the Mn$^{3+}$ ions in the $B$ sites. We ascribe this phenomenon to the high symmetry of the oxygen sites and to the absence of oxygen defects, two characteristics of quadruple perovskites not found in simple ones, which prevent the release of the Bi$^{3+}$-induced strain through distortions or disorder. This result demonstrates the possibility of achieving a large magnetoelectric coupling mediated by internal strain in proper ferroelectrics and suggests a novel concept of internal strain engineering for multiferroics design.
\end{abstract}

\maketitle


\section{\label{intro}Introduction}
The possibility of mutually controlling electric polarization and magnetization in multiferroic materials may lead to new device concepts for electronics. The challenge is to design materials with an effective magnetoelectric coupling suitable for applications. A first strategy is to achieve a coupling between polar and magnetic ions in the same compound \cite{eer06,che07,wan09}. This is typically the case of Bi-based compounds, such as the perovskite system Bi$B$O$_3$, where the 6$s^2$ lone pair of the Bi$^{3+}$ ion forms an electric dipole, while the $B$ site is occupied by a magnetic ion, such as Mn$^{3+}$ or Fe$^{3+}$ \cite{eer06,che07,wan09,kim03,smo82,kub90}. Despite intense research, modest magnetoelectric couplings have been hitherto reported in these compounds; this has been explained by the fact that the mere coexistence of polar and magnetic ions does not necessarily lead to their coupling \cite{eer06,che07,wan09}. Indeed, in these compounds, the ferroelectric order develops independent of the magnetic one (\textit{proper} ferroelectricity). Recently, the focus has thus been moved to \textit{improper} ferroelectrics, where ferroelectricity is induced by magnetism. Among the manganese oxides relevant here, notable are the $R$MnO$_3$ \cite{hua97,kat01} and $R$Mn$_2$O$_5$ \cite{sai95,cha04,bla05} systems ($R$=Y or rare earth), where a sizable tunability of the polarization by applied fields has been reported \cite{kim03b,hur04}. Various magnetoelectric coupling mechanisms have been considered, such as the spin-current interaction in spiral magnetic structures or the exchange striction in frustrated magnets \cite{mos06,jia07,hu08,rad09,moc10}, and the debate has attracted a great deal of interest.

In the absence of a firm theoretical framework, the possibility of a large coupling between polar and magnetic ions remains open and new hints are provided by recent experiments. Notable is a striking enhancement of magnetoelectric response in epitaxial BiFeO$_3$ films, where strain drives a rhombohedral to pseudo-tetragonal phase transition \cite{wan03,zha06}. Although the mechanism of this enhancement remains unclear, this result suggests that the magnetoelectric coupling in proper ferroelectrics is sensitive to strain and structural distortions \cite{inf10}. To clarify this point would be important for multiferroics design. This is a challenging task for the Bi$B$O$_3$ system owing to the complex structure-property relationships, e.g. polymorphism \cite{die11,mon05,bel07}, modulated structure \cite{mon05} and sensitivity to oxygen vacancies \cite{sun08,bel09,ede05}.

In this paper, we address the above point in a simpler system. By neutron powder diffraction, we have studied the effect of the polar Bi$^{3+}$ ion on the magnetic ordering in \bimno~ \cite{mez09,oka10}, the counterpart with \textit{quadruple} perovskite structure of the \textit{simple} perovskite BiMnO$_3$. Quadruple perovskites are described by the general formula \QP~ \cite{des67,mar73,pro04} and we shall restrict to the single-valent manganese oxides $A$Mn$_3$Mn$_4$O$_{12}$, where $A$ is a three-valent ion, such as La \cite{pro09,cab10,oka09}, Pr \cite{mez09b}, Bi \cite{mez09,oka10}, and Mn$^{3+}$ occupies both $A'$ and $B$ sites. The structure (see Fig. 1) consists of the same pseudocubic network of corner-sharing $B$O$_6$ octahedra characteristic of simple perovskites $AB$O$_3$, such as the aforementioned BiMnO$_3$ and TbMnO$_3$ multiferroics. Our motivation is that, in spite of this similarity, quadruple perovskites display a simpler distortion pattern and no oxygen vacancies \cite{pro04,pro09,sai10}. Our results show a striking correlation between the internal strain field induced by the polar Bi$^{3+}$ ions and the magnetic ordering of the Mn$^{3+}$ ions in \bimno. This conclusion provides a hint to achieve an effective magnetoelectric coupling between polar and magnetic ions in perovskite-like compounds.  

\begin{figure}[b]
\includegraphics[width=\columnwidth]{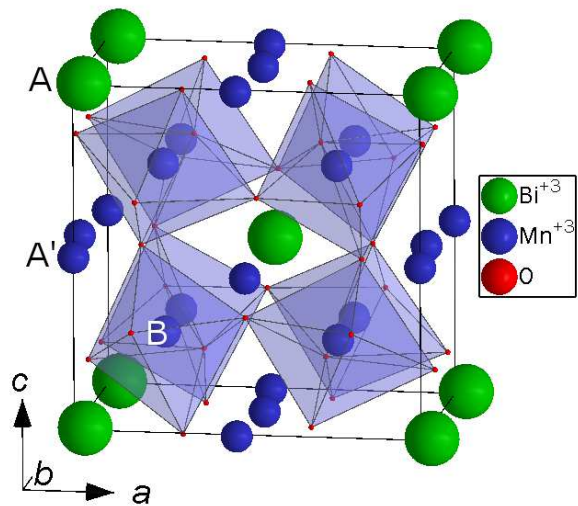}
\caption{\label{fig:structure} (Color online) Nuclear unit cell of the quadruple perovskite structure of \bimno. Note that the Mn$^{3+}$ ions occupy two distinct $A'$ and $B$ Jahn-Teller sites with square and octahedral coordination, respectively.}
\end{figure}

\subsection{\label{qp}Structural characteristics of quadruple perovskites $A$Mn$_3$Mn$_4$O$_{12}$}

As discussed in previous papers \cite{pro04,pro09}, the above favorable characteristics arise from the following features of quadruple perovskites. 1. Their structure is obtained by doubling the cubic unit cell axis of simple $AB$O$_3$ perovskites; the new unit cell therefore contains two distinct $A$ and $A'$ sites at the corners and middle of the cube edges, respectively. This doubling is caused by a large buckling of the $B$O$_6$ octahedra driven by the Jahn-Teller (JT) distortion of the $A'$ site. As a result, the Mn-O-Mn bond angle is drastically reduced to $\sim 135^\circ$, an unusually small value as compared to the $\sim 160^\circ$ value of simple perovskites, such as LaMnO$_3$, which typically display orthoferrite-type distortions \cite{rod98}. 2. The large buckling lowers the coordination number of the $A'$ site from twelve to four, which prevents oxygen vacancies, as these would destabilize the structure. Indeed, no evidence of vacancies has been hitherto reported in any quadruple perovskite \cite{mez09,pro04,pro09,mez09b,sai10}. No extra oxygen atoms can be hosted either within the structure owing to its high-density \cite{cab10,mez09b}. 3. The simple distortions arise from the high symmetry of the cubic $Im$\={3} phase, where all oxygen atoms occupy a unique high symmetry $m..$ site, which only permits a rigid tilt of the $B$O$_6$ octahedra \cite{pro09}. The JT distortion of the octahedra requires a cubic $Im$\={3} to monoclinic $I2/m$ phase transition typically realized at low temperature.

\subsection{\label{la_bi}Single-valent Mn$^{3+}$ quadruple perovskites \lamno~ and \bimno~}
In view of the above considerations, quadruple perovskites are a model system to investigate the coupling between Bi$^{3+}$ and Mn$^{3+}$ ions and \bimno~ is an obvious choice. It was previously reported \cite{mez09,oka10} that the nuclear structure of this compound is similar to that of the isovalent \lamno~ \cite{pro09,oka09}, where La$^{3+}$ instead of Bi$^{3+}$ occupies the $A$ site. A structural study on single crystals \cite{mez09} shows that the main difference between the two structures is the center of symmetry breaking in the former, which is described by the $Im$, instead of $I2/m$, space group. This study confirms that this breaking is caused by the polar properties of the Bi$^{3+}$ ion, which leads to a sizable spontaneous polarization $\sim 7 \mu$C cm$^{-2}$ at room temperature. In addition, \bimno~ displays an AFM transition at $T_N$ = 55 K concomitant to an anomaly of the dielectric constant, which suggests a coupling between AFM order and polarization. Interestingly, \lamno~ exhibits a similar AFM transition in the same temperature range $T_N$=78 K but no dielectric anomaly \cite{pro09,oka09}. The above results motivate the present study of the magnetic structure of \bimno.

\section{\label{exp}Experimental methods}
\subsubsection{\label{synthesis}High-pressure synthesis of \bimno~ powders}
Powder samples of single-phase \bimno~ were synthesized under high pressure in a multi-anvil apparatus at 4 GPa and 1000 $^\circ$C for one hour as discussed in detail elsewhere \cite{mez09}. After the high pressure and high temperature synthesis, the samples were quenched down to room temperature in less than 1 minute in order to stabilize the high pressure phase. In order to check phase purity, the as-prepared samples were first analyzed using a commercial Siemens X-ray powder diffractometer equipped with a Cu K$_{\alpha}$ radiation.

\subsubsection{\label{neutrons}Neutron powder diffraction measurements}
The nuclear and magnetic structures were studied at 300 K and in the 1.5-100 K range by means of neutron powder diffraction data collected at the 3T2 and G4-1 diffractometers, respectively, of the Laboratoire L\'eon Brillouin in Saclay, France. The 3T2 instrument is a high-resolution diffractometer with wavelength $\lambda$=1.225 \AA, whilst G4-1 is a high intensity diffractometer optimized for magnetic structure determination with $\lambda$=2.43 \AA. Both nuclear and magnetic structures were refined using the FullProf Suite package \cite{rod93}.

\section{\label{res}Results}
\subsection{\label{nstruct}Nuclear structure of \bimno~ at room temperature}

We first consider the room temperature diffractogram of Fig. 2. The data analysis shows that the as-prepared samples are 96\% pure or better with Mn$_3$O$_4$ (haussmannite) as main impurity \cite{cha86}. In agreement with Ref. \cite{mez09}, the room temperature structure was successfully refined in the \textit{noncentrosymmetric} space group $Im$, with $a$=7.545(1) \AA, $b$=7.362(1) \AA, $c$=7.536(1) \AA, $\beta=91.178(2)^\circ$. In Fig. 2, we compare the observed and calculated diffractograms. The refined structural parameters are reported in Table I, where we note a good agreement between observed and calculated intensities, with $R_p$ = 4.21\%, $wR_p$ = 5.57\%, and $\chi^2$= 3.67. The salient features of the refined structure are seen by considering the simplified cubic $Im$\={3} structure. In this structure, the MnO$_6$ octahedra are regular with the Mn$^{3+}$ ions ($B$ sites) in the center of each octant of the unit cell; the square-coordinated Mn$^{3+}$ ions ($A'$ sites) are located in the middle of the cell axes; the Bi$^{3+}$ ions ($A$ sites) occupy the cell corners and center; all twenty four oxygen atoms occupy the same site with high $m..$ symmetry. In the real $Im$ structure of Table I, all sites split into sites with lower symmetry and lower multiplicity. For instance, the centrosymmetric eightfold B site of Mn with \textit{cubic} .\={3}. symmetry splits into two \textit{noncentrosymmetric} fourfold sites Mn4 and Mn5 with translational .1. symmetry only. The displacements of the Mn$^{3+}$ and O$^{2-}$ ions allowed by the $Im$ symmetry lead to a macroscopic polarization of the crystal, \textbf{P}, approximately oriented along the $ac$-diagonal, in accord with the quasi isotropic properties of the $ac$-plane. On the other hand, the local dipoles (or microscopic polarization) of the MnO$_6$ octahedra have no constraints because the symmetry of the Mn4 and Mn5 sites is purely translational. Thus, these dipoles have a non-zero component along the $b$-axis. In Table II we report an estimate of the dipoles computed using a point-like charge model.

\begin{figure}[b]
\includegraphics[width=\columnwidth]{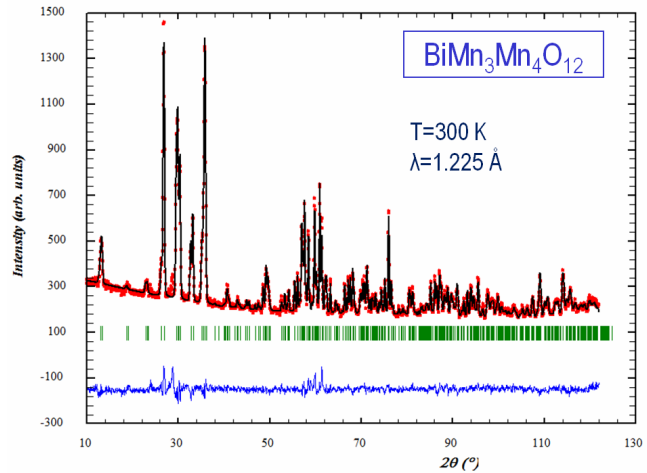}
\caption{\label{fig:refnucl} (Color online) Rietveld refinement of the neutron powder diffraction data of \bimno~ at room temperature. The wavelength used is $\lambda$=1.225 \AA. Circles and continuous line indicate the observed and calculated profiles, respectively. Vertical bars mark the positions of the calculated Bragg reflections. The continuous line at the bottom is the difference between observed and calculated patterns. The refined structural parameters are reported in Table I.}
\end{figure}

\begin{table*}
\label{tab:bimnostruct}
\caption{Refined structural parameters of \bimno~ at 300 K within the $Im$ space group. Numbers in parentheses indicate statistical uncertainty. Atomic coordinates $x$, $y$ and $z$ are given in reduced lattice units. Statistical indicators of the refinement are reported at the bottom.}
\begin{ruledtabular}
\begin{tabular}{ccccccc}

\multicolumn{7}{l} 
{Space Group: $Im$; $a$=7.545(1) \AA, $b$=7.362(1) \AA, $c$=7.536(1) \AA, $\beta=91.178(2)^{\circ}$}\\
\hline

Atom & Wyckoff position & Site symmetry & $x$ & $y$ & $z$ & $B_{iso}$ (\AA$^2$) \\

\hline
Bi        & 2a & $m$ & 0.032(3) & 0       &   0.006(3)  & \footnote{ Anisotropic thermal factors ($\times 10^4$):
$\beta_{11}$=11(8), $\beta_{2}$=215(25), $\beta_{33}$=53(16), $\beta_{12}$=0, $\beta_{13}$=10(8), $\beta_{23}$=0.}\\
Mn1($A'$) & 2a & $m$ & 0        & 0       &   0.5       & 0.18(5) \\
Mn2($A'$) & 2a & $m$ & 0.494(4) & 0       &   0.454(3)  & 0.18(5) \\
Mn3($A'$) & 2a & $m$ & 0.469(3) & 0       &  -0.005(4)  & 0.18(5) \\
Mn4($B$)  & 4b & 1   & 0.237(3) & 0.737(2)&   0.256(3)  & 0.18(5) \\
Mn5($B$)  & 4b & 1   & 0.236(3) & 0.762(2)&   0.747(3)  & 0.18(5) \\
O1        & 2a & $m$ & 0.326(3) & 0.5     &   0.178(3)  & 0.3(1)  \\
O2        & 2a & $m$ & 0.177(3) & 0       &   0.664(3)  & 1.0(1)  \\
O3        & 2a & $m$ & 0.165(4) & 0       &   0.286(3)  & 0.5(1)  \\
O4        & 2a & $m$ & 0.800(4) & 0       &   0.301(4)  & 0.8(1)  \\
O5        & 4b & 1   & 0.477(3) & 0.818(2)&   0.315(4)  & 0.9(1)  \\
O6        & 4b & 1   & 0.298(3) & 0.833(2)&  -0.026(3)  & 1.0(1)  \\
O7        & 4b & 1   & 0.010(3) & 0.686(2)&   0.162(3)  & 1.0(1)  \\
O8        & 4b & 1   & 0.665(3) & 0.193(2)&   0.001(3)  & 0.8(1)  \\
\hline

\multicolumn{7}{l} 
{Reliability factors with all non-excluded points for pattern:}\\
\multicolumn{7}{l} 
{$R$-factors (not corrected for background): $R_p$=4.21, $wR_p$=5.57, $R_{exp}$=2.93, $\chi^2$=3.62}\\
\multicolumn{7}{l} 
{Conventional Rietveld $R$-factors for pattern: $R_p$=16.2, $wR_{p}$=16.2, $R_{exp}$=8.50, $\chi^2$=3.62}\\
\multicolumn{7}{l} 
{Global user-weigthed $\chi^2$ (Bragg contribution)=3.67}\\

\end{tabular}
\end{ruledtabular}
\end{table*}

\begin{table}

\label{tab:dipoles}
\caption{Components along the $a$-, $b$- and $c$-axis of the local electric dipole \textbf{d} on the MnO$_6$ octahedra estimated using a point-like charge model for the Mn$^{3+}$ and O$^{2-}$ ions. Units are $q_e$\AA, where $q_e$ denotes the electron charge.}

\begin{tabular}{cccc}

\hline\hline
Mn site & $d_a$  & $d_b$  & $d_c$ \\
\hline
Mn4     & -0.072 &  0.399 & 0.453 \\
Mn5     & -0.018 & -0.207 & 0.246 \\
\hline\hline

\end{tabular}

\end{table}

\begin{table}
\label{tab:mndist}
\caption{Mn-O distances in \AA~ for the octahedral Mn$^{3+}$ $B$ sites. Numbers in parentheses indicate statistical uncertainty.}

\begin{tabular}{ccc}

\hline\hline
Mn site & O site & Mn-O distance \\

\hline

Mn4     & O1     & 1.96(2) \\
Mn4     & O7     & 1.88(3) \\
Mn4     & O8     & 1.96(3) \\
Mn4     & O6     & 2.29(3) \\
Mn4     & O5     & 1.95(3) \\
Mn4     & O3     & 2.02(2) \\
Mn5     & O8     & 1.98(3) \\
Mn5     & O6     & 1.84(3) \\
Mn5     & O2     & 1.91(2) \\
Mn5     & O4     & 2.03(2) \\
Mn5     & O5     & 2.11(3) \\
Mn5     & O7     & 2.21(3) \\
\hline\hline

\end{tabular}

\end{table}

In Table III we report the six Mn-O bond lengths of the MnO$_6$ octahedra and note that they vary within a broad 1.840-2.295 \AA~ range. Hence, the description of the crystal field in terms of $e_g$ $x^2-y^2$ and 3$z^2-r^2$ orbitals is no longer appropriate, which prevents a straightforward application of the Goodenogh-Kanamori-Anderson (GKA) rules \cite{goo63} to estimate the sign and the strength of the exchange interaction between neighboring Mn$^{3+}$ ions. Another reason of inapplicability is the large buckling of the MnO$_6$ octahedra. Indeed, the Mn-O-Mn bond angle of $\sim 135^\circ$ falls in between the opposite limits of $90^\circ$ and $180^\circ$ considered by the above rules.

\subsection{\label{mstruct}Low-temperature diffraction data: magnetic orderings of \bimno~}
We now consider the evolution of the diffraction data in the 1.5-100 K range (see Fig. 3). These data indicate that the symmetry of the nuclear structure is $Im$ at all temperatures. At 55 and 25 K, two new sets of diffraction peaks show up, which are attributed to the two AFM orderings probed by previous magnetization measurements \cite{mez09}. The indexing of the two sets is consistent with two distinct \textbf{k}$_1$=($1/2$,0,$1/2$) and \textbf{k}$_2$=(0,0,0) propagation vectors, respectively. This is similar to the case of the isovalent \lamno, which exhibits an independent AFM ordering of the Mn$^{3+}$ $B$ and $A'$ sublattices at comparable temperatures $T_{N,B}$=78 K and $T_{N,A'}$=21 K, respectively. Since the exchange interaction between Mn$^{3+}$ ions is expected to be similar in \bimno, for \bimno~ as well we propose the same scenario of two independent AFM orderings for the $B$ and $A'$ sublattices, with $T_{N,B}$=55 K and $T_{N,A'}$=25 K. The question of the absence of coupling between the two magnetic sublattices in quadruple perovskites has been discussed elsewhere \cite{pro04,sai10}.

\begin{figure}[b]
\includegraphics[width=\columnwidth]{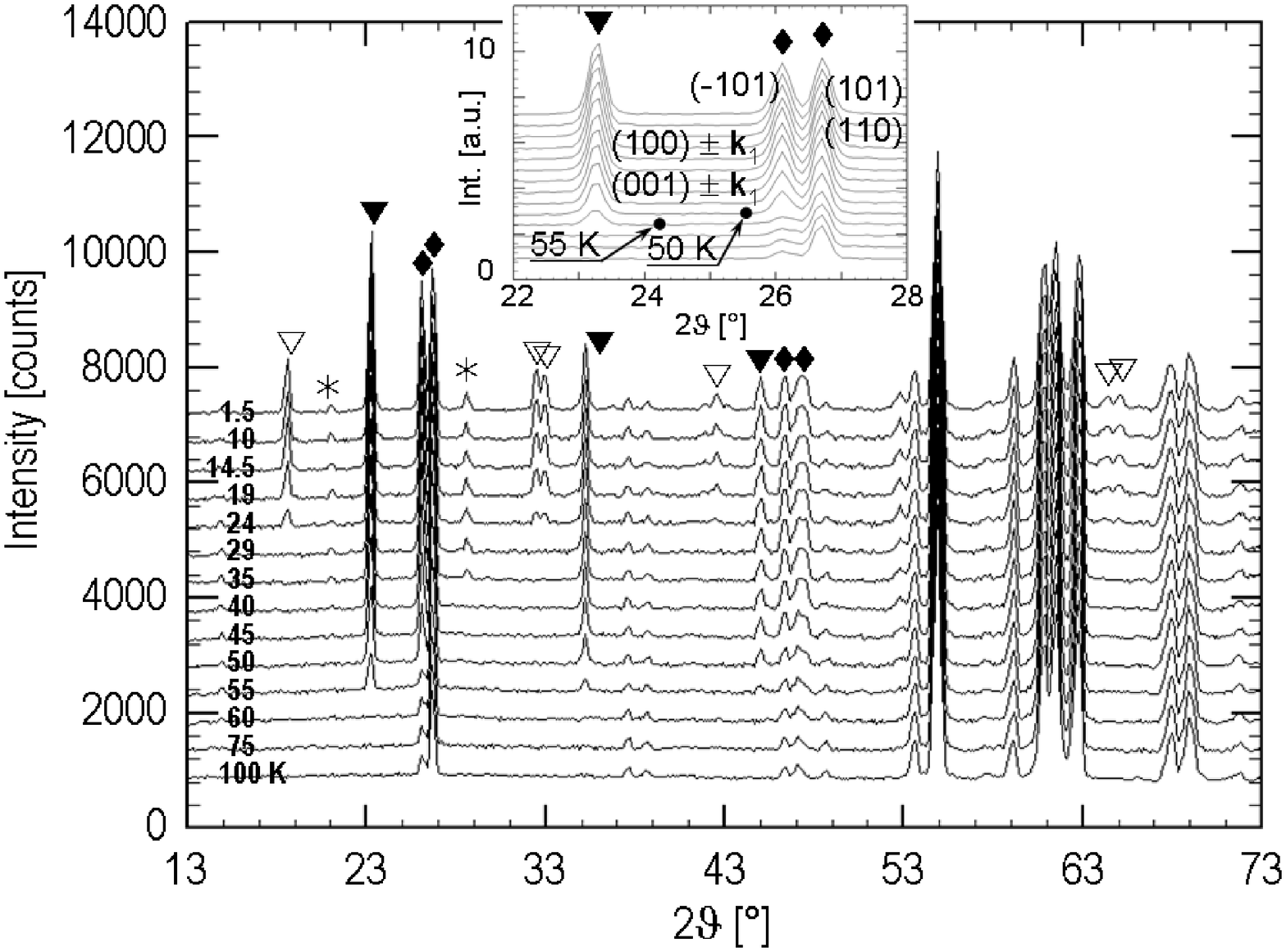}
\caption{\label{fig:diffr_T} Temperature dependence of the powder neutron diffraction patterns of \bimno. The wavelength used is $\lambda$=2.43 \AA. Three magnetic transitions are visible at 55 ($\blacktriangledown$), 50 ($\blacklozenge$) and 25 K ($\triangledown$). The first two ones refer to the ordering of the magnetic $B$-sublattice, whilst the third one refers to the $A'$-sublattice. The ($\blacktriangledown$) symbol labels the peaks associated with the \textbf{k}$_1$=($1/2$,0,$1/2$) propagation vector, whilst the ($\blacklozenge$) symbol labels the Bragg peaks whose intensities exhibit a jump at 50 K and are therefore associated with the \textbf{k}$_2$=(0,0,0) propagation vector. The ($\ast$) symbol labels the peaks of the antiferromagnetic transition of the Mn$_3$O$_4$ impurity. Inset: detail of the behavior of the (\={1}01) and (110)-(101) doublet Bragg peaks and of the (100) $\pm$ \textbf{k}$_1$-(001) $\pm$ \textbf{k}$_1$ magnetic doublet peak.}
\end{figure}

Fig. 3 shows that the AFM transition at 55 K is characterized by an unusual feature. The appearance of the magnetic peaks indexed by \textbf{k}$_1$ is followed at slightly lower temperature (50 K) by a sudden increase of intensity of certain Bragg peaks. Since the magnetic response of the system is AFM, this increase indicates that the AFM structure of the $B$ sublattice contains a \textit{uniform} component described by the \textbf{k}$_2$ propagation vector. On the other hand, the AFM ordering of the $A'$ sublattice is completely described by \textbf{k}$_2$ and corresponds to the loss of the $I$ centering (anti-body centering) of the lattice.

\subsection{\label{ref_mstruct}Refinement of the magnetic structure of \bimno~}

To solve the magnetic structures of both sublattices, we determined the basis functions of the irreducible representations of the $Im$ space group for the relevant propagation vectors and Wyckoff sites of the Mn$^{3+}$ ions (see Table IV). The $\Gamma_{mag}=3\Gamma_1 \oplus 3\Gamma_2$ decomposition into irreducible representations is obtained for the two Mn4 and Mn5 sites of the $B$ sublattice and for both \textbf{k}$_1$ and \textbf{k}$_2$ vectors, whilst the $\Gamma_{mag}=\Gamma_1 \oplus 2\Gamma_2$ decomposition is obtained for the three Mn1, Mn2 and Mn3 sites of the $A'$ sublattice and for the only relevant \textbf{k}$_2$ vector. Thanks to the above analysis, we performed simulated annealing refinements \cite{kir83} of the diffraction data with the following constraints: 1. the total magnetic moment has been set to zero for both $A'$ and $B$ sublattices, as the magnetization data of Ref. \cite{mez09} give evidence of an AFM response; 2. we have assumed that the candidate structures follow either the $\Gamma_1$ or $\Gamma_2$ representation, not an admixture of the two. For the refinements, the two 1.5 and 35 K data set were used in order to solve separately the structures of the $A'$ and $B$ sublattices. Indeed, the ordering of the $B$ sublattice is complete at 35 K, as apparent from the temperature dependence of the magnetic peak intensities in Fig. 3 and of the lattice parameters in Fig. 5. Also, 35 K is sufficiently higher than the ordering temperature $T_{N,A'}$ = 25 K of the $A'$ sublattice, so the structure of the latter can be neglected.

\begin{table}
\label{tab:basis}
\caption{Basis functions, $\psi$, for the axial vectors associated with the irreducible representations $\Gamma_1$ and $\Gamma_2$ for the two different $B$ and $A'$ sites of the magnetic Mn$^{3+}$ ions and for the two different propagation vectors \textbf{k}$_1$ and \textbf{k}$_2$ that describe the magnetic structure of \bimno.}

\begin{ruledtabular}
\begin{tabular}{|ccc|ccc|}
\multicolumn{6}{|c|}{4b Wyckoff site (octahedral $B$ site), \textbf{k}$_1$=($1/2$ 0 $1/2$)}\\
\hline
$\Gamma_1$&($x$ $y$ $z$)&($x$ $-y$ $z$)&$\Gamma_2$&($x$ $y$ $z$)&($x$ $-y$ $z$)\\
\hline
$\psi_1$ &  1  0   0  & -1  0   0  & $\psi_1$ &  1   0   0  &  1  0  0\\
$\psi_2$ &  0  1   0  &  0  1   0  & $\psi_2$ &  0   1   0  &  0 -1  0\\
$\psi_3$ &  0  0   1  &  0  0  -1  & $\psi_3$ &  0   0   1  &  0  0  1\\
\hline\hline

\multicolumn{6}{|c|}{4b Wyckoff site (octahedral $B$ site), \textbf{k}$_2$=(0 0 0)}\\
\hline
$\psi_1$ &  1  0   0  & -1  0   0  & $\psi_1$ &  1   0   0  &  1  0  0\\
$\psi_2$ &  0  1   0  &  0  1   0  & $\psi_2$ &  0   1   0  &  0 -1  0\\
$\psi_3$ &  0  0   1  &  0  0  -1  & $\psi_3$ &  0   0   1  &  0  0  1\\
\hline\hline

\multicolumn{6}{|c|}{2a Wyckoff site (square-coordinated $A'$ site), \textbf{k}$_2$=(0 0 0)}\\
\hline
$\psi_1$  &  0  1   0  &            &          &             & \\
          &            &            & $\psi_2$ &  1   0   0  & \\
          &            &            & $\psi_3$ &  0   0   1  & \\

\end{tabular}
\end{ruledtabular}
\end{table}

\begin{figure}[b]
\includegraphics[width=\columnwidth]{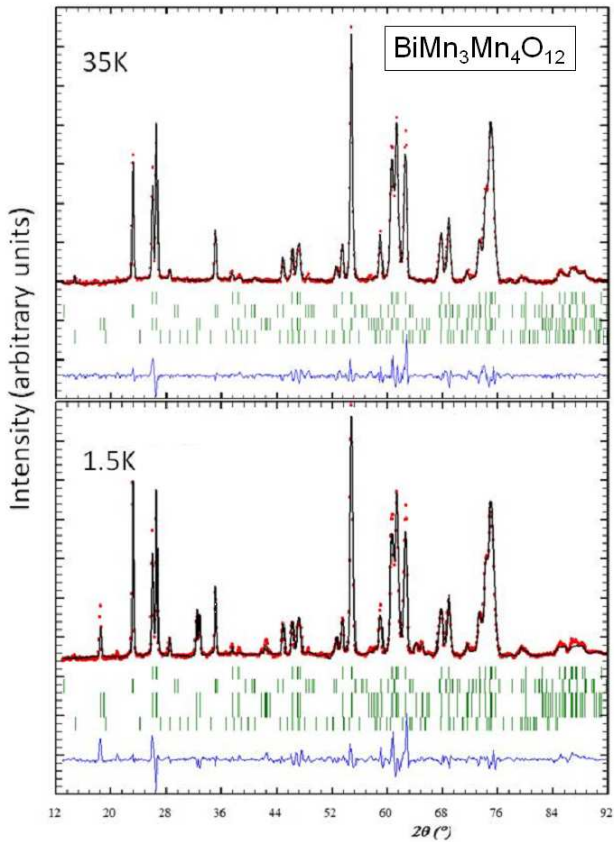}
\caption{\label{fig:refmag} (Color online) Rietveld refinement of the powder neutron diffraction data of the preceding figure. Top panel: refinement of the 35 K data. Circles and continuous line represent the observed and calculated diffraction patterns, respectively. Vertical bars mark the calculated position of the nuclear peaks (first row), the magnetic peaks of the \textbf{k}$_1$-component of the Mn($B$) structure appearing at 55 K (second row), the magnetic peaks of the \textbf{k}$_2$-component of the Mn($B$) structure appearing at 50 K (third row) and the nuclear peaks of the Mn$_3$O$_4$ impurity (fourth row). The continuous line at the bottom is the difference between observed and calculated patterns. Bottom panel: the same as for the top panel for the 1.5 K diffraction pattern. The additional row of ticks indicates the magnetic peaks of the Mn($A'$) structure.}
\end{figure}

\subsubsection{\label{mstructB}Magnetic structure of the $B$-sublattice}
As to the $B$ sublattice, the magnetic structural models described by $\Gamma_1$ yielded the best statistical results. The result of the refinement is shown in Fig. 4 and Table V, whilst the structure is illustrated in Fig. 5. We outline the characteristics of this structure by analyzing separately the two \textbf{k}$_1$- and \textbf{k}$_2$-components with moments parallel and perpendicular to the $ac$-plane, respectively. The \textbf{k}$_1$-component consists of an $E$-type AFM structure with moments in the $ac$ plane $M_{B,ac}$=1.37 $\mu_B$. This structure can be viewed as zig-zag chains along the $a$- and $c$-axis with alternate antiparallel and parallel moments. We recall that \lamno~ displays a different $C$-type structure made of ferromagnetically (FM) coupled AFM $ac$-planes \cite{pro09}. It is remarkable that two compounds with similar nuclear structures and isoelectronic properties exhibit radically different magnetic orders. This is a first indication of the role played by the polar Bi$^{3+}$ ion on the magnetic ordering. As to the magnetic structure of the \textbf{k}$_2$-component, Table V and Fig. 5 show that the moments are oriented along the $b$-axis and that they are as large as $M_{B,b}$=1 $\mu_B$. In view of the discussion below, it is noted that the orientation depends on the crystallographic site, i.e. the moments of the Mn4 (Mn5) sites are parallel (antiparallel) to the $b$-axis. In summary, the moments $\textbf{M}_{B}$ of the $B$ sites result from an unusual superposition of two components $\textbf{M}_{B,ac}$ and $\textbf{M}_{B,b}$ with distinct \textbf{k}$_1$ and \textbf{k}$_2$ propagation vectors. Hence, the magnetic structure can be viewed as a large and uniform modulation $\pm \textbf{M}_{B,b}$ of an $E$-type structure. The total moment resulting from this superposition is $M_{B}$=1.70 $\mu_B$, a much smaller value than the value of 4.0 $\mu_B$ expected for the high spin state ($S$=2) of the Mn$^{3+}$ ion. This discrepancy was reported for \lamno~ as well \cite{pro09}, thus confirming the inadequacy of a purely ionic model for the electronic states of manganese oxides with quadruple perovskite structure.

\begin{figure}[b]
\includegraphics[width=\columnwidth]{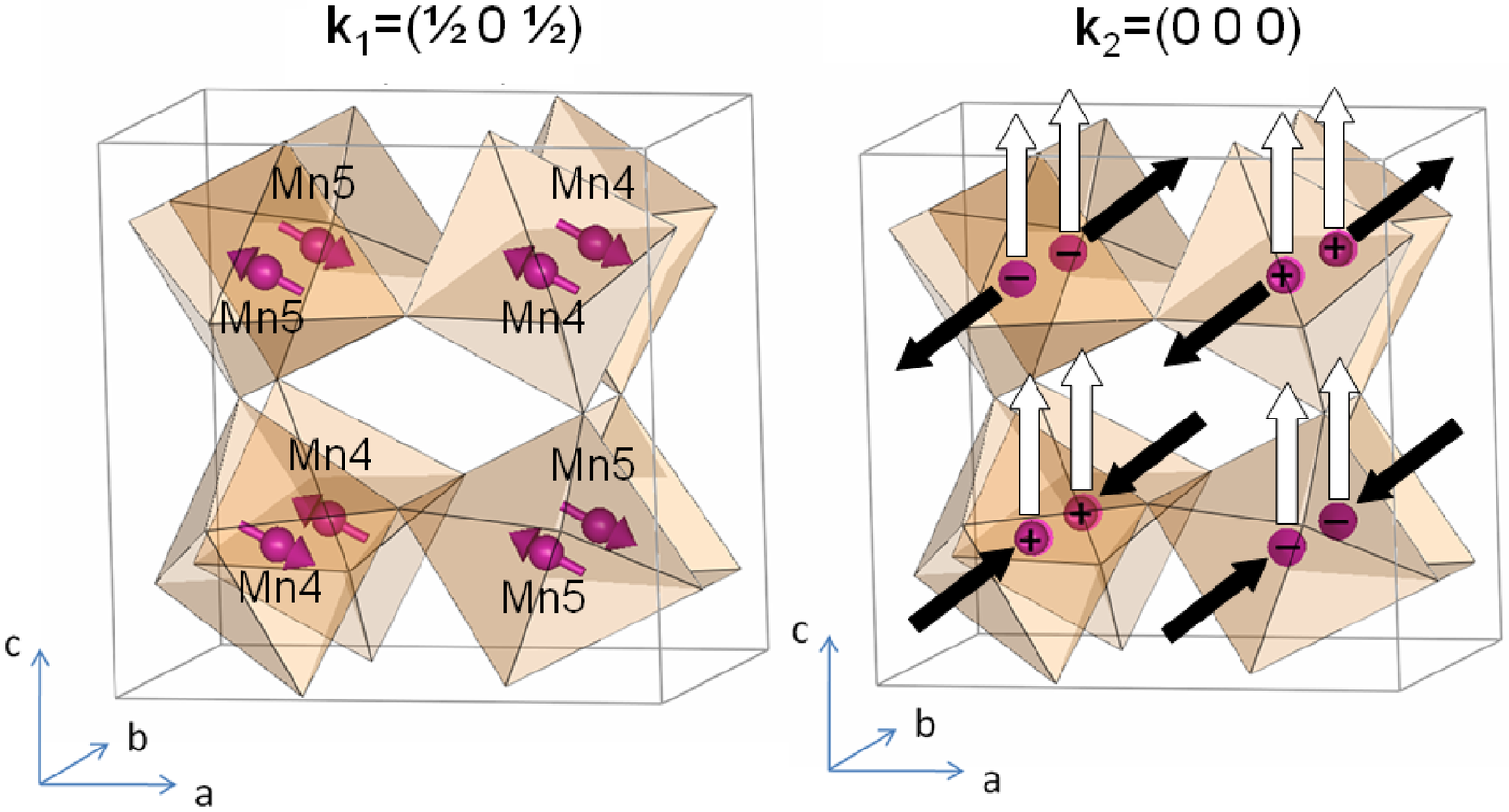}
\caption{\label{fig:magB} (Color online) Schematic representation of the two components with propagation vectors \textbf{k}$_1$=($1/2$,0,$1/2$) (left panel) and \textbf{k}$_2$=(0,0,0) (right panel) which form the AFM structure of the Mn$^{3+}$ $B$-sublattice of \bimno. The purple balls indicate Mn$^{3+}$ ions. The purple arrows indicate the moments of the Mn$^{3+}$ ions, whilst the $\pm$ signs indicate the sign of the $b$-component of the moments whilst the black (white) arrows (not in scale) indicate the $b$- ($c$-) components of the electric dipoles formed by the MnO$_6$ octahedra (the small $a$-component is not shown for simplicity). The magnetic moments of the structure in the left panel are reversed in the adjacent nuclear cells along the $a$- and $c$-axis, as the structure is described by the \textbf{k}$_1$ propagation vector. This is not the case for the structure of the right panel, as this is described by \textbf{k}$_2$=(0,0,0).}
\end{figure}

\begin{figure}[b]
\includegraphics[width=\columnwidth]{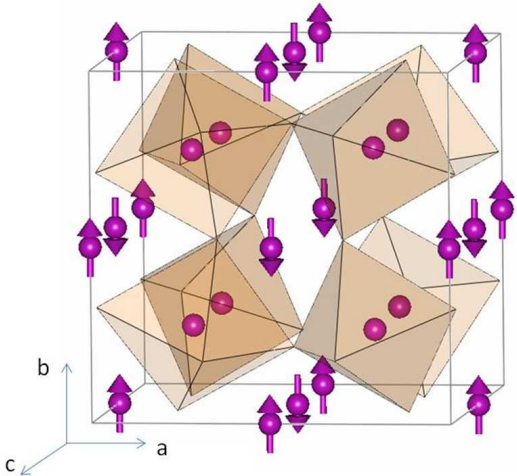}
\caption{\label{fig:magA} The same as in the preceding Figure for the Mn$^{3+}$ $A'$-sublattice. }
\end{figure}  

\begin{table*}
\label{tab:mstruct}
\caption{Magnetic structures of the $B$ and $A'$ sublattices of \bimno~ at 1.5 K. The structures are obtained after refinement of the diffraction data of Figure 4. The nuclear unit cell parameters are: $a$=7.537(1) \AA; $b$=7.354(1) \AA; $c$=7.534(1) \AA; $\alpha$, $\gamma$=90$^{\circ}$; $\beta$= 91.213(1)$^{\circ}$. Magnetic moments are given in Cartesian coordinates and in Bohr magneton $\mu_B$ units. Numbers in parentheses indicate statistical uncertainty. Approximate atom coordinates are given. The magnetic structure of both sublattices is described by the $\Gamma_1$ representation, i.e. the moments ($M_x$,$M_y$,$M_z$) transform as ($-M_x$,$M_y$,$-M_z$) upon application of the mirror operation ($x$,$y$,$z$) $\rightarrow$ ($x$,$-y$,$z$) on atom coordinates (see text and Table IV).}

\begin{ruledtabular}
\begin{tabular}{lcccc}
\multicolumn{5}{c}{Mn$^{3+}$ $B$-sublattice ($T_{N,B}$=55 K) ; propagation vector \textbf{k}$_1$=($1/2$ 0 $1/2$)}\\
\hline

Atom coordinates       & $M$     & $M_x$   & $M_y$ & $M_z$ \\

\hline
Mn4 ($1/4$ $3/4$ $1/4$)& 1.37(1) &-1.22(4) & 0    & 0.60(6) \\
Mn4 ($3/4$ $1/4$ $3/4$)& 1.37(1) &-1.22(4) & 0    & 0.60(6) \\
Mn5 ($1/4$ $3/4$ $3/4$)& 1.37(1) & 1.22(4) & 0    &-0.60(6) \\
Mn5 ($3/4$ $1/4$ $1/4$)& 1.37(1) &-1.22(4) & 0    & 0.60(6) \\

\hline\hline

\multicolumn{5}{c}{Mn$^{3+}$ $B$-sublattice ($T_{N,B}$=50 K); propagation vector \textbf{k}$_2$=(0 0 0)}\\
\hline

Mn4 ($1/4$ $3/4$ $1/4$)& 0.99(2) & 0   & 0.99(2) & 0 \\
Mn4 ($3/4$ $1/4$ $3/4$)& 0.99(2) & 0   & 0.99(2) & 0 \\
Mn5 ($1/4$ $3/4$ $3/4$)& 0.99(2) & 0   &-0.99(2) & 0 \\
Mn5 ($3/4$ $1/4$ $1/4$)& 0.99(2) & 0   &-0.99(2) & 0 \\

\hline\hline

\multicolumn{5}{c}{Mn$^{3+}$ $A'$-sublattice ($T_{N,A'}$=25 K); propagation vector \textbf{k}$_2$=(0 0 0)}\\
\hline
Mn1 (0 0 $1/2$)     & 1.22(1) & 0 & 1.22(1) &0 \\
Mn1 ($1/2$ $1/2$ 0) & 1.22(1) & 0 &-1.22(1) &0 \\
Mn2 ($1/2$ 0 $1/2$) & 1.22(1) & 0 &-1.22(1) &0 \\
Mn2 (0 $1/2$ 0)     & 1.22(1) & 0 & 1.22(1) &0 \\
Mn3 ($1/2$ 0 0)     & 1.22(1) & 0 & 1.22(1) &0 \\
Mn3 (0 $1/2$ $1/2$) & 1.22(1) & 0 &-1.22(1) &0 \\

\end{tabular}
\end{ruledtabular}
\end{table*}

\subsubsection{\label{mstructA}Magnetic structure of the $A'$-sublattice}
Also in the case of the AFM structure of the $A'$ sublattice, the best structural model is described by the $\Gamma_1$ representation. The magnetic moments and the structural model are reported in Table V and Fig. 6, respectively. The structure consists of an anti-body centered AFM ordering with a moment $M_{A'}$=1.22 $\mu_B$ per site. The structure turns out to be identical to that of \namno \cite{pro04} but different from that of \lamno \cite{pro09}. This confirms the key role played by Bi$^{3+}$ in altering the stability of competing magnetic orderings in these two isoelectronic and almost isostructural compounds. Similarly to the previous case of the $B$-site, also the moment per $A'$ site turns out to be much smaller than that expected according to a purely ionic model.  

\section{\label{PvsM}Correlation between lattice polarization and magnetic ordering}
We should now provide a physical interpretation of the magnetic orderings observed. Our point concerns the symmetry properties of the Bi$^{3+}$-induced lattice polarization and of the AFM structure of the Mn$^{3+}$ $B$-sublattice reported in Fig. 5 and in Table V. We show below that these properties point to a large magnetoelectric coupling in \bimno. We first note that the order parameters of the polarization and of the $b$-component, $M_b$, of the magnetic moments \textbf{M} are both described by the \textbf{k}$_2$ propagation vector. This suggests that $M_b$ is a \textit{uniform} modulation of the $E$-type structure induced by the noncentrosymmetric displacements of the Mn$^{3+}$ and O$^{2-}$ ions via a magnetoelastic coupling. To show the validity of this scenario, we recall that the dominant contribution of this coupling to the Landau free energy is of the form $f = -\beta^{ijkl} \epsilon_{ij} M_k M_l$, where $\beta$ is a fourth-rank coupling tensor, $\epsilon$ is the strain tensor and \textit{i,j,k,l} are the indices of the $a$, $b$ and $c$-axis \cite{lan84}. In our case, the coupling arises from the \textit{internal} strain associated with the dipoles of the MnO$_6$ octahedra (see Table II and Fig. 5). Indeed, this field \textit{uniformly} alters the electrostatic potential and the spin-orbit interaction of the Mn$^{3+}$ ion as well as the exchange integrals of the Mn-O-Mn bonds. This problem was extensively treated by Callen and Callen \cite{cal63}, who considered both the external strain and the internal one caused by the atomic displacements. We recall that, by symmetry, the spins can couple only to strain modes symmetric under spatial inversion. In our case, due to the mirror $ac$ plane of our pseudocubic $Im$ symmetry, the only relevant mode is the $\Gamma_1$ mode symmetric under the $y \rightarrow -y$ inversion. This observation confirms the validity of our scenario of magnetoelastic coupling between $M_b$ and the polarization-induced strain, since the $\Gamma_1$ representation also describes $M_b$. The reason is that the free energy contains the fully symmetric representation to be obtained from the direct product of the two representations of the strain and spin basis functions \cite{lan84,cal63}. Thus, the two representations must be equivalent, as in our case. The existence of a coupling is confirmed by the fact that an internal strain along the $b$-axis does indeed exist, as the dipoles of the MnO$_6$ octahedra have a sizable component along the $b$-axis. Owing to the mirror $ac$-plane, the atomic displacements associated with the dipoles create a modulation of the Mn-Mn distance along the $b$-axis (i.e. an internal strain $\epsilon_{bb}$), the short (long) distance being as small (large) as 3.490 (3.858) \AA.

\begin{figure}[b]
\includegraphics[width=\columnwidth]{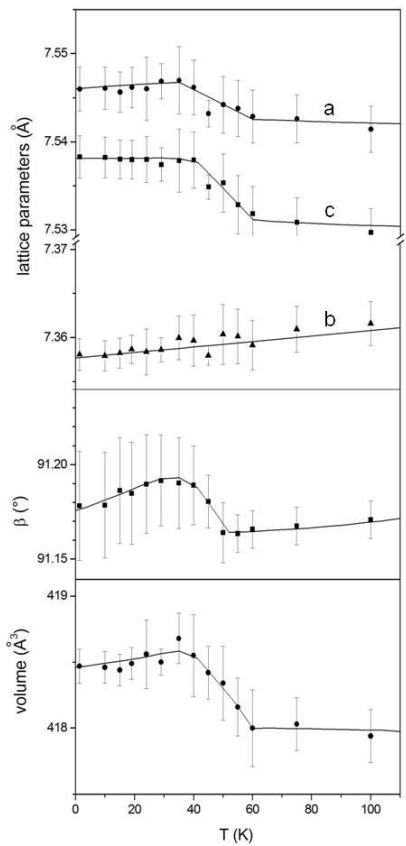}
\caption{\label{fig:lattice_par} Temperature dependence of the lattice parameters of \bimno. The data are obtained from the refinement of the powder neutron diffractograms of Fig. 2. Note the large magnetostriction along the $a$- and $c$-axis at the antiferromagnetic ordering at 55 K of the Mn$^{3+}$ ions in the octahedral $B$-sublattice (see text).}
\end{figure}

The above considerations also account qualitatively for the stability of the $M_b$ structure. Indeed, the moments are bound to be parallel along the $b$-axis because of the $\Gamma_1$ symmetry of the strain field. In addition, the moments on the Mn4 and Mn5 sites tend to be antiparallel as the exchange interaction is predominantly AFM within the $ac$-plane \cite{pro09}. Both characteristics are indeed found in our experiment. A quantitative account of the stability of the $M_b$ structure goes beyond the scope of this work; it may be provided by \textit{ab initio} calculations in the local spin density approximation that include the on-site repulsion between $d$-electrons or \textit{ad hoc} Hamiltonians that include the charge, orbital and spin degrees of freedom. These methods have been applied to \lamno~ \cite{liu10}, $R$MnO$_3$ \cite{fen05,ser06,mal09} and $R$Mn$_2$O$_5$ \cite{bar11} among others, and an application may be envisaged in our case as well.

Our picture of magnetoelastic coupling along the $b$-axis also accounts for the temperature-dependence of the unit cell parameters reported in Fig. 7. One notes that the AFM ordering of the $B$ sites at 55 K is accompanied by a sizable expansion of both $a$- and $c$-axes, whilst the $b$-axis displays no anomaly. This indicates a magnetostriction only in the $ac$-plane, which is explained as follows. The variations of the unit cell parameters reflect the \textit{external} strain induced by the \textit{macroscopic} polarization, \textbf{P}. As mentioned before, in \bimno, \textbf{P} is approximately oriented along the $ac$ diagonal, while its $b$-component is identically zero. Therefore, we expect a magnetostriction only along this diagonal, in agreement with Fig. 7. Similar arguments account for the absence of magnetostriction at the AFM transition of the Mn$^{3+}$ ions at the $A'$ sites at $T_{N,A'}$=25 K. Indeed, the $b$-component of the dipole at this site is identically zero because of the $m$ symmetry of these sites. Thus, the only symmetric internal strain mode $\Gamma_1$ available for the magnetoelastic coupling vanishes in this case.

\section{\label{concl}Conclusions}
In conclusion, we found that the magnetic order of the Mn$^{3+}$ ions in the quadruple perovskite \bimno~ is strongly modulated by the internal strain field induced by the polar Bi$^{3+}$ ion. This observation suggests a mechanism of magnetoelectric coupling between Bi$^{3+}$ and Mn$^{3+}$ ions driven by the magnetoelastic coupling due to the internal strain field, in agreement with a previous observation of giant magnetoelastic effect in the multiferroic $R$MnO$_3$ system \cite{lee08}. Our analysis indicates that this mechanism is effective in \bimno~ thanks to two features of quadruple perovskites: a) the high symmetry of the oxygen atoms surrounding the Bi$^{3+}$ ion, which limits structural distortions, thus preventing the release of the strain; b) the absence of oxygen disorder. Further experimental and theoretical work is needed to test the tunability of the magnetic order of \bimno~ by an electric field, a prerequisite for applications.

In the light of the above considerations, our result demonstrates that a large magnetoelectric coupling between magnetic and polar ions can be achieved in proper ferroelectrics. The coupling mechanism which accounts for our result suggests that site symmetry is an effective control parameter of the internal strain responsible for the coupling. This conclusion suggests a novel concept of internal strain engineering for the effective design of multiferroics.

\bibliography{manga}

\end{document}